\documentclass[useAMS,usenatbib]{mn2e}

\usepackage{times,graphicx,amssymb}
\newcommand{\etal}{et al.}
\newcommand{\mnras}{MNRAS}
\newcommand{\apj}{ApJ}

\newcommand{\apjl}{ApJL}
\newcommand{\apjs}{ApJS}
\newcommand{\physrep}{Phys.~Rep.}
\newcommand{\aap}{A\&A}

\newcommand{\prd}{PhRvD}
\newcommand{\araa}{ARA\&A}

\title[First lensing measurements of SZ-detected clusters]
  {First lensing measurements of SZ-detected clusters}
\author[R.\  N.\ McInnes \etal]
  {Rachel N. McInnes,$^1$\thanks{rnm@roe.ac.uk}
  Felipe Menanteau,$^2$ Alan F. Heavens,$^1$ John P. Hughes,$^2$
  \newauthor 
Raul~Jimenez,$^3$ Richard~Massey,$^1$ Patrick~Simon,$^1$  and Andy Taylor$^1$\\
  $^1$Scottish Universities Physics Alliance (SUPA), Institute for Astronomy, University of Edinburgh, Blackford Hill, Edinburgh, EH9 3HJ, U.K.\\
  $^2$Rutgers University, Department of Physics \& Astronomy, 136 Frelinghuysen Road, Piscataway, NJ 08854, USA\\
  $^3$ICREA \& ICE(CSIC)-IEEC, UAB campus, Bellaterra 08193, Spain }
\date{}

\pubyear{2009}

\def\LaTeX{L\kern-.36em\raise.3ex\hbox{a}\kern-.15em
    T\kern-.1667em\lower.7ex\hbox{E}\kern-.125emX}

\begin{document}

\label{firstpage}

\maketitle

\begin{abstract}
We present the first lensing mass measurements of Sunyaev-Zel'dovich (SZ) selected clusters. Using optical imaging from the Southern Cosmology Survey (SCS), we present weak lensing masses for three clusters selected by their SZ emission in the South Pole Telescope survey (SPT).  We confirm that the SZ selection procedure is successful in detecting mass concentrations. We also study the weak lensing signals from 38 optically-selected clusters in $\simeq 8$ square degrees of the SCS survey. We fit Navarro, Frenk and White (NFW) profiles and find that the SZ clusters have amongst the largest masses, as high as $5 \times 10^{14}M_\odot$. Using the best fit masses for all the clusters, we analytically calculate the expected SZ integrated $Y$ parameter, which we find to be consistent with the SPT observations. 
\end{abstract}

\section{Introduction}

Galaxy clusters are the largest gravitationally bound objects in the Universe and can be used as cosmological probes because their formation and evolution rate are sensitive to different cosmological parameters \citep[e.g.][]{evrard1989, Haiman2001, Allen2004}.   The abundance of galaxy clusters as a function of mass $N(m,z)$ at high redshift $z$ is particularly sensitive to different cosmological models. To probe cosmology and dark energy we must observe galaxy clusters at high redshift and obtain mass estimates for them. 

Observations of the Sunyaev-Zel'dovich (SZ) effect \citep*{SZ1981} are a powerful way to probe galaxy clusters by detecting the hot cluster gas \citep{Birkinshaw99}.  SZ-detected clusters are in principle particularly powerful as they can be seen to high redshifts.  The intensity of the SZ effect summed over the entire cluster closely tracks the mass of the cluster \citep{Motl2005}. X-ray or SZ effect mass estimates are based on simplified assumptions such as a hydrostatic equilibrium for the cluster gas. It is becoming increasingly apparent, however, that we cannot fully model the complex gas physics in clusters within a simple framework. Nonetheless, it will be very challenging to calibrate cluster masses at high redshift, so in order to use SZ observations to probe cluster properties and cosmological models it is important to understand the relationship between mass and SZ observables in lower redshift systems. Gravitational lensing facilitates the calibration of the SZ observables to obtain accurate masses for SZ detections \citep{LewisKing,SealfonJimenez}. A large area in the southern sky is currently being surveyed in SZ by the Atacama Cosmology Telescope (ACT) and the South Pole Telescope (SPT).

Gravitational lensing is dependent only on the projected mass distribution of the lens and so it is possible to study the mass distribution independent of its form, including the distribution of dark matter. Gravitational lensing causes small ($\sim$ a few $\%$) changes to the shape of individual galaxies, which can be used to reconstruct the mass distribution in the region \citep{KS93}; for a review see \citet{Munshi2008}.   In this paper we present measurements of weak lensing masses for clusters which were, for the first time, detected blind by their SZ decrement \citep{SPTSZ}. We also include mass measurements from 24 optically-detected clusters. This paper is structured as follows: in \S\ref{sec:observations} we present the data and discuss the image processing; \S\ref{sec:WL} describes the gravitational lensing methods used; the mass measurements are presented  in \S\ref{sec:results} and we compare with measurements from other techniques and calculate the $Y$ parameters. Throughout this paper we assume a flat cosmology with $\Omega_{m} =0.27 $, $\Omega_{\Lambda} = 0.73$ and $H_{0} = 100h$ km~s$^{-1}$Mpc$^{-1}$ with $h = 0.71$.

\section{Observations}
\label{sec:observations}

\label{sec:data}
We use publicly-available data from the Blanco Cosmology Survey - a National Optical Astronomy Observatory Large Survey Project observing 60 nights over 4 years on the Blanco 4m telescope at the Cerro Tololo InterAmerican Observatory in Chile. The Mosaic II camera is being used for a deep, four-band optical (\emph{griz}) survey of two 50 deg$^{2}$ patches of the southern sky. Two areas of southern sky have been targeted, centred on $23^{h}00^{m}$, $-55^{\circ}12^{m}$ and $05^{h}30^{m}$, $-52^{\circ}47^{m}$. These fields lie within a larger area of the southern sky which ACT and SPT plan to survey. The paper is based on observations taken in 2005, with the exception of 2 clusters from 2006 data. The seeing varies between $0.81''$ and $1.09''$ with a mean of $0.89''$. The image reduction was carried out using the Rutgers Southern Cosmology Pipeline (flat field correction, CCD calibration, removal of saturated star bleed-trails, and bad pixel masks). Next the images were aligned, stacked and median combined using SWarp \citep{Bertin:2006}; an astrometric solution was found by matching stars to sources in the US Naval Observatory Catalog. Additional masks were made to remove saturated stars, satellite trails and other blemishes in the image, removing 8\% in total.  For more information see \cite{BCS1}. Note that we use AB magnitudes throughout.  To calculate photometric redshifts (photo-$z$s), multi-band SExtractor \citep{sex} \emph{g,r,i,z} isophotal magnitudes were used to find redshift probability distributions of each object. This was done using the {\sc bpz} code \citep{BPZ}, see also \cite{BCS1}.  We focused on 4 clusters found in SZ \citep{SPTSZ} and also 38 optical clusters selected from $\simeq 8$ square degrees \citep{BCS1}. Of the 38 clusters, we found non-zero mass estimates for 24 clusters, and upper limits for the remaining clusters. We do not consider 5 of the clusters as they are in regions only observed in a single exposure.

\section{Weak Lensing Analysis}
\label{sec:WL}
We used the \emph{i}-band data, 3 co-added exposures of 450s each, for our shear analysis, and measured galaxy shapes using the \citet{KSB95} (KSB) method.  Our pipeline is based on \cite{bre} and labelled ``MB'' in \cite{STEP1}, but has been automated to process rapidly the large SCS data set.  The method deviates from ``MB'' in automated star/galaxy separation and in Point Spread Function (PSF) interpolation. We tested the pipeline against simulated images from the Shear TEsting Programme (STEP) \citep[see][]{STEP1,STEP2}. Our method underestimates shear, but consistently throughout a wide range of observing conditions, which is henceforth compensated for in our shear measurements by applying a calibration factor of $1/(0.82\pm 0.05)$, similar to ``MB". 

As in ``MB" our pipeline locates galaxies via the {\sc imcat}\footnote{{http://www.ifa.hawaii.edu/$\sim$kaiser/imcat}} {\sc hfindpeaks} algorithm and measures their quadrupole shape moments using a Gaussian weight function of width $4r_g$, where $r_{g}$ is the size of the best-fitting Gaussian. The pipeline fits the shear polarizability factor $\frac{1}{2}{\mathrm{Tr}}(P^\gamma)$ as a function of galaxy size. We excluded galaxies smaller than $1.1$ times the measured seeing, and those with $S/N<5$, leaving $9$ per square arcminute to \emph{i}$\sim$23. The median redshift is $0.65$ \citep{Coil}; 69\% of galaxies in our catalogue also had photo-$z$s, which have a consistent $z$ distribution \citep{BCS1}.  Departing from the ``MB'' pipeline, we performed star/galaxy separation via the automated {\sc theli} algorithm \citep{THELI}, and automated the removal of galaxies with abnormally large values of shear polarizability, smear polarizability or ellipticity, which was present in the ``MB" method but labor intensive and slow. The observed shapes of galaxies were finally corrected for the blurring effect of the PSF. We measured the PSF using the $0.5$ unsaturated stars per square arcminute with $22.0>i>18.1$ (S/N $55$ to $1670$). The PSF ellipticity is  $0.035\pm 0.019$, where the error is the standard deviation throughout the survey. Optical effects and temporal variation of the atmosphere and telescope between dithered exposures produce patterns in PSF size and ellipticity. This variation was fitted as a sum of polynomials (of order $4$ in the $x$ and $y$ directions) plus sums of sines and cosines (of orders up to $4$ in $x$ and $8$ in $y$). These choices give a small r.m.s. residual $|e|$ of $0.0092$.

We estimate masses by fitting an NFW profile \citep*{NFW} to the average tangential shear  from {\sc imcat's etprofile} routine. The concentration index of the halo was taken as a function of mass and cluster redshift \citep{dolag2004}. For a robust treatment of missing regions of data we use a Wiener-filtered mass reconstruction method \citep{HuKeeton2002}. To address the mass-sheet degeneracy the average surface mass density is set to zero over the whole field of view (half a square degree).   For the SPT clusters the peaks of the surface mass density in the Wiener-filtered maps were, where present, assumed to be at the centre of the cluster, otherwise (SPT~0547-5345) the luminosity-weighted centre was used. The latter procedure was followed for the optically selected clusters. 
Our Wiener-filtered reconstruction was done as follows: the two-point correlation function of the lensing convergence, used as prior for the Wiener reconstruction, was estimated from the shear-shear correlation \citep{bs2001} $\xi_+(\theta)$ in the data itself. In order to have a smooth prior, we fitted the measured $\xi_+$ with $\xi_+(\theta)=a/(1+b\,\theta)$ where $a$ and $b$ are constants determined by the fit. To obtain S/N maps of the lensing maps we divided the maps by the r.m.s. of 500 noise realisations, which were generated by randomly rotating the ellipticities of the sources followed by a Wiener reconstruction. The S/N maps were then used to confirm the cluster centres found in the mass maps.


\section{Results}
\label{sec:results}

\begin{figure}
\centering
 \includegraphics[width=65mm]{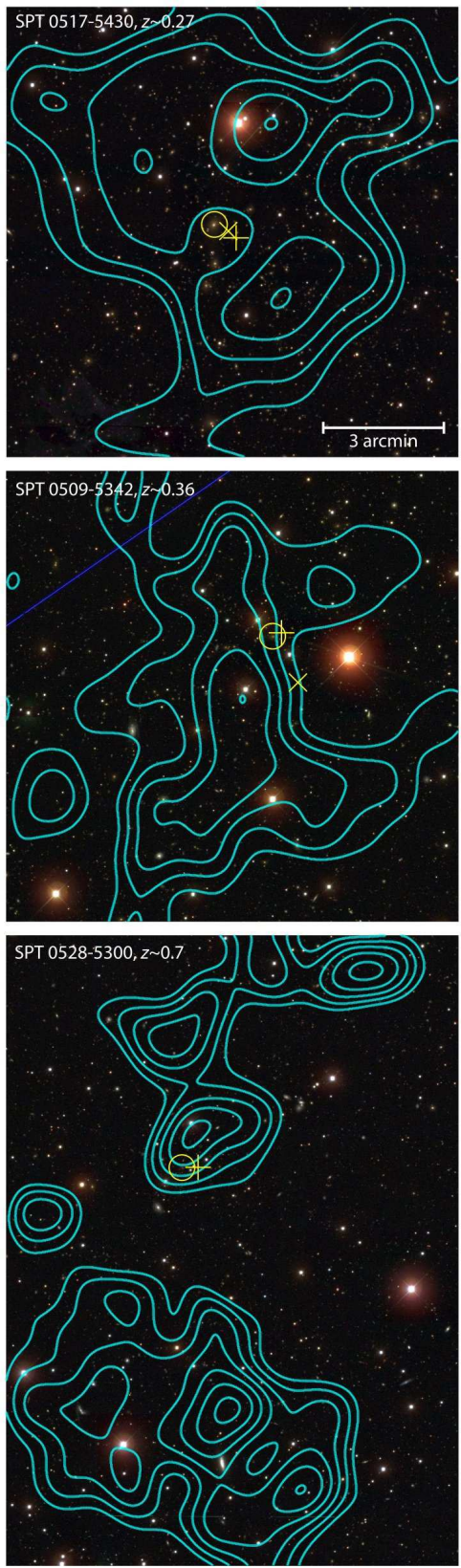}
 \caption{Lensing mass reconstructions for SZ-detected clusters. Contours show the projected lensing convergence (mass distribution), at $1\%$, $1.5\%$, $2\%$, ... $4\%$. A circle indicates the BCG in each optically selected cluster, an $\times$ indicates a peak in X-ray emission, and a + denotes the SZ peak.}
  \label{fig:SPT123}
\end{figure}

Wiener-filtered mass reconstructions for three of the SPT SZ-detected clusters are shown in Figure~\ref{fig:SPT123}. The fourth SZ cluster at redshift $z\sim 0.88$ was not detected. Our mass maps of clusters SPT0517-5430, SPT0509-5342, and SPT0528-5300 have peak S/N=3.2, 3.0 and 2.5 respectively. The mass reconstruction of SPT0528$-$5300, at $z\sim0.7$, is shown in the lower panel of Fig.~\ref{fig:SPT123}. There is a second peak (S/N = 3.5) which may warrant further investigation; it is $6.5'$ from the brightest cluster galaxy (BCG). It is at $05^{h}27^{m}46.2^{s}$, $-53^{h}07^{m}57.9^{s}$ and we estimate it to be $8.12^{+5.03}_{-7.34}\times10^{14} M_{\sun}$. On examination of \citet[][Fig.1]{SPTSZ}, a small decrement in the SZ appears to be present. There is no statistically significant difference between the redshift distribution of galaxies in a $3'$ radius around this location compared to the redshift distribution of the surrounding $0.4$ square degree region. 

\begin{figure}
\centering
 \includegraphics[width=60mm]{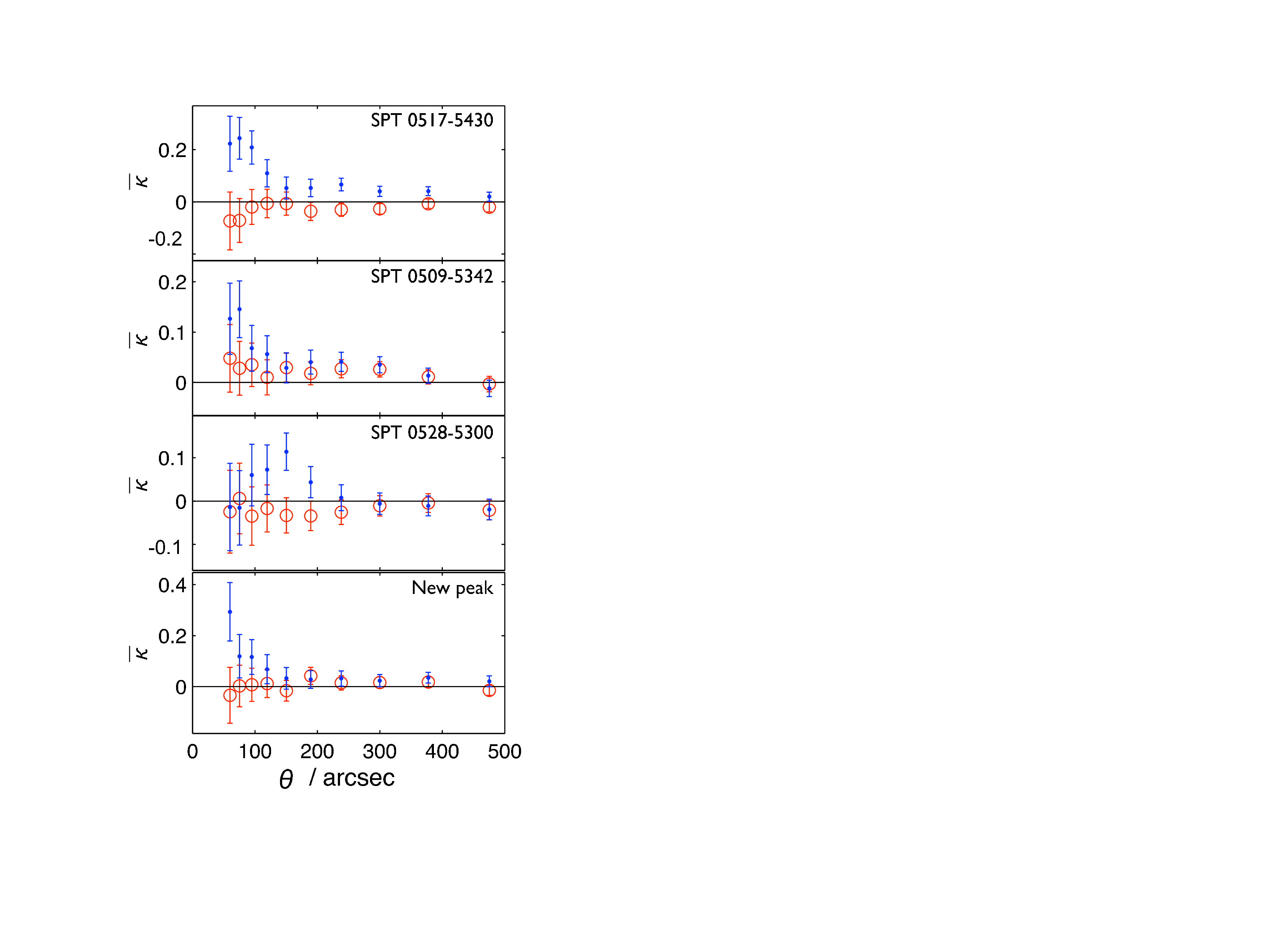}
 \caption{Average convergence within an aperture, $\bar\kappa(\theta)$ for the three detected SPT clusters plus the new offset peak found near SPT0528-5300. The red (empty circle) shows the B-mode systematic error.}
  \label{fig:mass_prof}
\end{figure}

\begin{figure}
\centering
 \includegraphics[width=80mm]{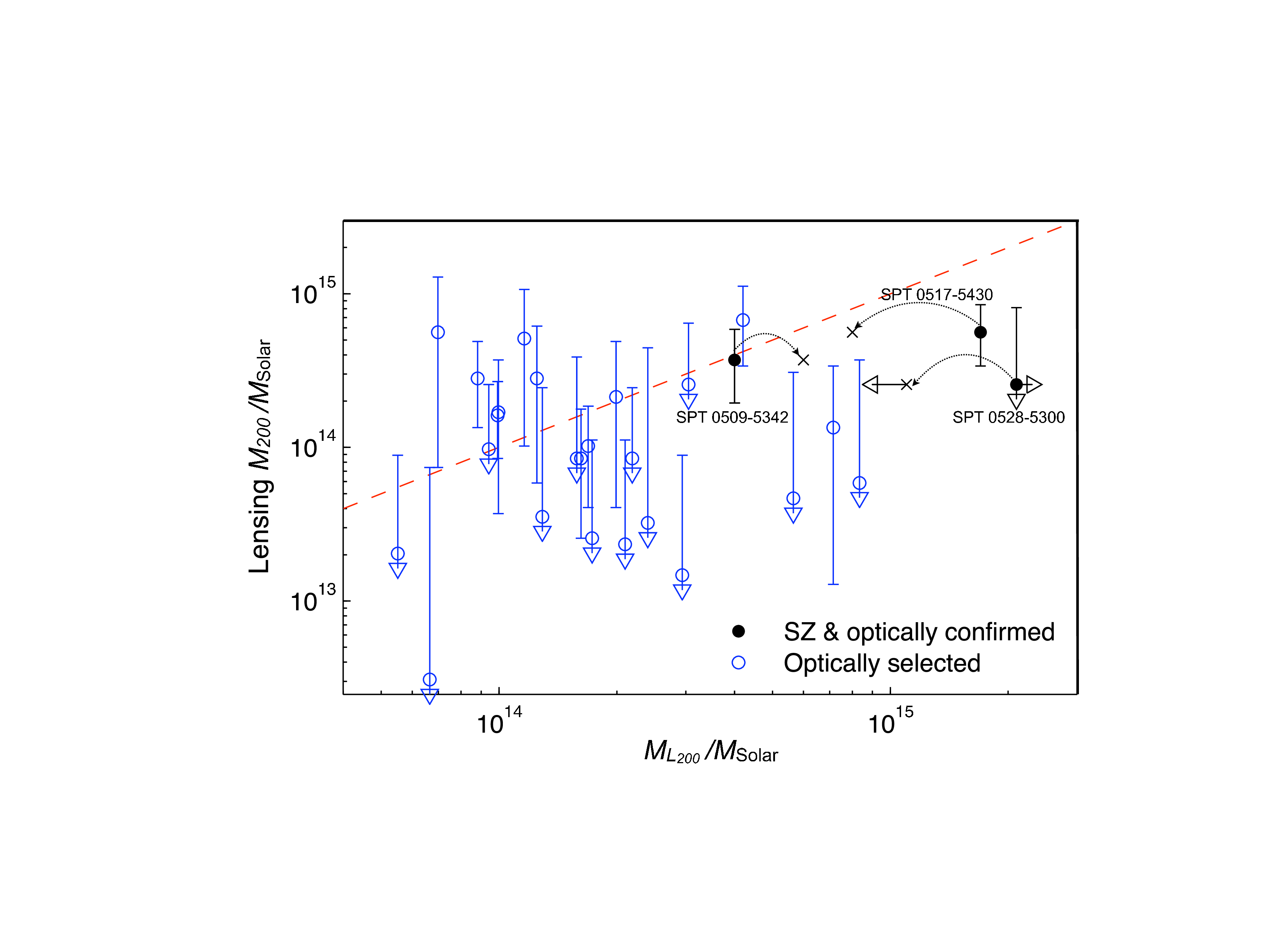}
 \caption{Lensing mass $M_{200}$ measurement against optical mass $M_{L_{200}}$ for SZ and optically selected clusters. Filled circles show the SZ-selected and optically confirmed clusters, while empty circles denote clusters observed optically. Lower error bars marked with a triangle signify that the lower error reaches zero. X-ray mass estimates are shown as an $\times$ on this plot for the 3 SZ-detected clusters, joined by an arc to the optical estimate. The dashed line is $M_{L_{200}}=M_{200}$ to guide the eye. The uncertainty in $M_{L_{200}}$ is estimated to be a factor 2 \citep{BCS1}.}
  \label{fig:mass_corr}
\end{figure}

Figure~\ref{fig:mass_prof} shows the average convergence in apertures for the three detected SPT clusters and for the offset peak. In red (empty circles) we show the B-mode which suggests that the shear catalogues are reasonably free of systematics. Figure~\ref{fig:mass_corr} compares lensing and optical mass estimates. We see that the SZ-discovered detections are amongst the most massive of the clusters $\ge 3 \times 10^{14}M_\odot$. $M_{L_{200}}$ is an estimate of the mass, calibrated by weak lensing measurements for SDSS clusters \citep{Reyes2008}, within a radius $r_{200}$ in which the number density is estimated to be $200/\Omega_{m}$ times the average galaxy number density. This is subject to uncertainties in bias, but is claimed to be an unbiased (to 5\%) estimate of the radius where the mass density is 200 times the critical density \citep{Johnston2007}. The uncertainty in $M_{L_{200}}$ is estimated to be a factor of 2 \citep{BCS1} due to the uncertainty in extrapolating the scaling relation to higher redshifts and uncertain  cluster membership. The correlation between the optical masses  \citep{BCS1, BCS2}  and the weak lensing masses gives some justification for using the optical luminosity as a mass proxy. Interestingly, the most discrepant of the SZ clusters has conflicting optical and X-ray mass estimates, suggesting that $M_{L_{200}}$ is overestimated.

We have also calculated the expected Compton $y$ parameter and its integral over solid angle, $Y \equiv \int {\rm d}\Omega\, y(\Omega)$ for the clusters. In Tables~\ref{tab1} and \ref{tab2} we show the expected temperature decrement in the Rayleigh-Jeans limit $\langle- \Delta T_{RJ}\rangle = 2T_{CMB} \langle y\rangle$, averaged within $r_{200}$. At 150GHz, the actual decrement is smaller by a factor $0.29$ \citep[][Fig. 2]{Carlstrom2002}. To compute $Y$ we assume that gas follows dark matter and obtain an analytic result. It ignores the gas history (cf. \citet{ReidSpergel2006}) but our simple model agrees, within $25\%$, with the empirical scaling relations of \citet{Motl2005} and \citet{Nagai2006},  and is supported by \citet{Atrio2008} who show from stacked SZ clusters that the baryon profile is consistent with NFW. $Y$ as a function of $M_{200}$ is

\begin{equation}\label{eq:eq1}
Y = \frac{\sigma_T}{6 \alpha m_e c^2}\left(\frac{\Omega_{b}}{\Omega_{m}}\right) \frac{1}{D_A^2}\frac{\delta_{200}^{1/3} M_{200}^{5/3}c_s (1+z) (G H_0)^{2/3}}{[\ln(1+c_s)-c_s/(1+c_s)]^2}
\end{equation}

\noindent where $D_A$ is the angular diameter distance, $M_{200}$ is the mass within an overdensity $\delta_{200} = 200$, $\alpha$ is the total pressure divided by the electron pressure. We assume that ions and electrons are in thermal equilibrium with a Helium mass fraction of 0.24, so $\alpha=1.93$. We also assume the concentration index 
\begin{equation}
c_s(M_{200})=9.59 (1+z)^{-1}(M_{200}/10^{14}h^{-1}M_\odot)^{-0.102}
\end{equation}
\citep{dolag2004}, but note that this is for a $\sigma_{8}=0.9$ cosmology, higher than current estimates \citep{WMAP5}. Values of $Y$ were predicted for clusters with mass detections, with statistical errors which are dominated by the error in $M^{Lens}_{200}$. 

With a uniform prior on masses, we record in Table~\ref{tab1} the most likely masses and asymmetric one-sigma errors for the SPT SZ clusters. X-ray masses from soft X-ray luminosity using the correlations in \citet{Reiprich2002} are shown in Table~\ref{tab1}. Predicted $Y$ and average $y$ within $r_{200}$ are also shown here, along with the cluster centre used. To the extent that we can estimate $\langle y \rangle$ from the published maps, our results seem to be about $50\%$ higher than observed, but within the errors expected from the mass determinations. Recent ACT SZ measurements \citep{Hincks2009} are in agreement with our predictions for the two low-redshift clusters which we can compare. Observations of SPT0517-5430 with XMM-Newton yield an X-ray mass within $r_{500}$ of $6.4\times 10^{14} M_{\odot}$ \citep{Zhang2006}. Table~\ref{tab2} shows mass measurements and predicted $Y$ and average $y$ within $r_{200}$ for the optically-detected clusters.

\begin{table*}
 \centering
 \begin{minipage}{180mm}
  \caption{Physical properties of SZ selected clusters in the SCS regions. Redshifts are the photometric redshift of the BCG, with $\pm1\sigma$ limits. The ID is based on the position of the BCG. The NFW fit is centred at either the peak in the mass map (map) where available, or the luminosity-weighted centre (lum.). The uncertainty in $M_{L_{200}}$ is estimated to be a factor 2 \citep{BCS1}. Values of $2T_{CMB}\langle y \rangle$ and $Y$ are predicted from $M^{Lens}_{200}$ using Eq.~\ref{eq:eq1}. Note that $\Delta T_{SZ}$ at 150GHz is 0.29 of the penultimate column value.}
  \begin{tabular}{rrrcrcclrr}
  \hline 
  ID & R.A. & Dec. & Centre & $z_{\rm photo}$ & $M_{L_{200}}$ & $M_{L_{\rm X}}$ & $M^{Lens}_{200}$ & $2T_{CMB}\langle y\rangle$ & $Y$\\
&   & Centre of NFW fit  & & & ($10^{14}M_{\sun}$) & ($10^{14}M_{\sun}$) & ($10^{14}M_{\sun}$)  & ($\mu K$) & ( $10^{-5}$ \\
& & & & & & & & & $arcmin^{2}$)\\
   \hline
SPT~0517-5430 & 05:16:27.3 & $-$54:27:39.4 & map & $0.27_{-0.02}^{+0.02}$  &  $17$     &   $8$ & $5.61_{-2.23}^{+2.88}$ & $280_{-114}^{+150}$ & $303_{-150}^{+170}$  \\
SPT~0509-5342 &  05:09:24.4 & $-$53:43:34.4 & map & $0.36_{-0.02}^{+0.02}$  & $4$     &   $6$ &$3.54_{-1.68}^{+2.07}$ & $219_{-106}^{+133}$ & $129_{-40}^{+40}$  \\
SPT~0528-5300 & 05:28:04.8 & $-$52:58:55.6 & map & $0.70_{-0.02}^{+0.03}$  & $\geq21$  &   $<11$ &  $2.94_{-2.94}^{+5.94} $ &$384_{-384}^{+839}$ & $63_{-63}^{+130}$ \\
SPT~0547-5345 & 05:46:41.1 & $-$53:44:52.1  & lum. & $0.88_{-0.04}^{+0.08}$  & $\geq4$  &   $10$ & - & - & - \\
\hline
\label{tab1}
\end{tabular}
\end{minipage}
\end{table*}

\begin{table*}
 \centering
 \begin{minipage}{180mm}
  \caption{As Table 1, but for optically selected clusters in the SCS regions.  The NFW fit is centred at the luminosity-weighted centre.}
  \begin{tabular}{rrrrclrr}
  \hline 
  ID & R.A. & Dec. & $z_{\rm photo}$ & $M_{L_{200}}$  & $M^{Lens}_{200}$ & $2T_{CMB}\langle y\rangle$ & $Y$\\
&   & Centre of NFW fit   & & ($10^{14}M_{\sun}$)  & ($10^{14}M_{\sun}$)  & ($\mu K$) & ( $10^{-5}$ \\
& &  & & & & & $arcmin^{2}$)\\
   \hline
SCSO~J232540.2-544430.9 & 23:25:32.2 & $-$54:44:21.5  & 0.10$\pm  0.02$ & 2.10&  $ 0.23_{-0.23}^{+0.89}$ & $7_{-7}^{+88}$ & $6_{-6}^{+75}$   \\ 
SCSO~J232230.9-541608.3 & 23:22:27.2 & $-$54:16:26.9  & 0.12$\pm 0.02$ & 1.62 &  $ 0.85_{-0.59}^{+0.92}$   &  $27_{-23}^{+64}$ & $39_{-33}^{+93}$ \\ 
SCSO~J233000.4-543707.7 & n/a & & 0.14$\pm  0.02$ & 1.19 & -& - & - \\ 
SCSO~J232419.6-552548.9  & 23:24:33.6 & $-$55:26:14.4  & 0.18$\pm  0.04 $ & 1.19  &$ <0.26 $ & - & -  \\ 
SCSO~J233106.9-555119.5 & 23:31:08.2 & $-$55:50:56.4  & 0.19$\pm 0.04$ & 0.55&  $ 0.20_{-0.20}^{+0.69} $ &  $8_{-8}^{+83}$ & $2_{-2}^{+19}$ \\ 
SCSO~J233252.9-561454.1 & 23:32:51.4 & $-$56:15:29.8  & 0.20$\pm 0.03$ &1.17 &  $ < 0.09$ & -& - \\ 
SCSO~J233215.5-544211.6 & 23:32:17.1 & $-$54:42:43.1  & 0.20$\pm 0.04$ & 1.69&  $ 1.02_{-0.61}^{+0.84}$& $40_{-32}^{+70}$ & $25_{-19}^{+42}$ \\ 
 SCSO~J233037.1-554338.8 &23:30:34.8& $-$55:43:41.5 & 0.20$\pm 0.04$ & 0.99&  $ 1.62_{-0.77}^{+1.07}$ & $65_{-31}^{+44}$ & $68_{-24}^{+50}$ \\ 
SCSO~J233550.6-552820.4 & 23:35:46.4 & $-$55:29:21.3  & 0.22$\pm 0.04 $ & 0.67 & $ 0.03_{-0.03}^{+0.71}$ &  $1_{-1}^{+250}$ & $0.1_{-0.1}^{+21}$  \\ 
SCSO~J232200.4-544459.7 &23:22:01.8 & $-$54:45:38.8  & 0.27$\pm 0.04$ & 1.73 &$ 0.26_{-0.26}^{+0.86}$ &  $12_{-12}^{+125}$ & $2_{-2}^{+18}$ \\ 
SCSO~J233522.6-553237.0 & 23:35:20.0 &$-$ 55:32:30.9  & 0.29$\pm 0.04 $ &  2.19 & $ 0.85_{-0.85}^{+1.60}$  & $42_{-42}^{+205}$ & $11_{-11}^{+55}$  \\ 
SCSO~J233807.5-560304.9 &23:38:07.7 & $-$56:02:55.0  & 0.30$\pm 0.04$ & 2.60& $< 0.64 $ &- & -  \\ 
 SCSO~J232956.0-560808.3 &23:29:55.8 & $-$56:08:28.2  & 0.32$\pm 0.04$ & 1.99 &$ 2.13_{-1.73}^{+2.75}$ & $117_{-110}^{+350}$ & $49_{-46}^{+75}$ \\ 
 SCSO~J232839.5-551353.8 &23:28:41.0 & $-$55:13:25.2 & 0.32$\pm 0.05$ &1.00 & $ 1.69_{-1.32}^{+2.01}$ & $92_{-85}^{+249}$ & $33_{-30}^{+88}$  \\ 
 SCSO~J232633.6-550111.5 &23:26:31.1 & $-$55:01:26.9 & 0.32$\pm 0.05$ & 2.81 &$< 0.48$ & -& -  \\ 
SCSO~J233753.8-561147.6 &23:37:57.1 & $-$56:12:05.8 & 0.33$\pm 0.04$ & 2.94 & $ 0.15_{-0.15}^{+0.74}$ & $8_{-8}^{+148}$ & $1_{-1}^{+9}$ \\ 
  SCSO~J232156.4-541428.8 & 23:21:55.5 & $-$54:14:20.0 & 0.33$\pm 0.04$ & 1.25 &$< 0.71$& -& - \\ 
SCSO~J233003.6-541426.7  & 23:30:06.3 & $-$54:13:58.9  & 0.33$\pm 0.04$ & 0.88 & $ 2.81_{-1.47}^{+2.07}$ & $160_{-113}^{+241}$ & $75_{-53}^{+114}$ \\ 
SCSO~J233231.4-540135.8 &   n/a & & 0.33$\pm 0.04$ & 1.67 & -& -& -  \\ 
SCSO~J233430.2-543647.5 &  n/a & & 0.35$ \pm 0.05$ & 3.59 & - & -& - \\
SCSO~J233110.6-555213.5 &23:31:08.4 & $-$55:51:38.3   & 0.39$\pm 0.05$ & 1.04 & $< 0.56$& -& - \\
SCSO~J233618.3-555440.3 & 23:32:13.8 & $-$55:54:16.4 & 0.49$\pm 0.03 $ & 0.94 &$0.97_{-0.97}^{+1.59} $& $78_{-78}^{+314}$ & $9_{-9}^{+36}$  \\
SCSO~J233706.3-541903.8 & 23:37:11.3  & $-$54:18:57.5 & 0.51$\pm 0.04$ & 1.58  & $0.85_{-0.85}^{+3.03}$& $71_{-71}^{+820}$ & $7_{-7}^{+80}$ \\
SCSO~J233816.9-555331.1 & 23:38:12.7 & $-$55:53:12.5 & 0.52$\pm 0.03$ & 1.29  & $0.35_{-0.35}^{+2.09}$& $29_{-29}^{+711}$ & $2_{-2}^{+37}$  \\
 SCSO~J233556.8-560602.3  & 23:35:55.6  & $-$56:05:50.5 & 0.52$\pm 0.03$ & 7.15  & $1.35_{-1.22}^{+2.03}$& $117_{-114}^{+423}$ & $15_{-15}^{+54}$ \\
 SCSO~J232619.8-552308.8 & 23:26:14.5  & $-$55:23:22.9 & 0.52$\pm 0.03$  & 1.25  & $2.81_{-2.22}^{+3.34}$ & $250_{-231}^{+672}$ & $52_{-48}^{+140}$  \\ 
SCSO~J233425.6-542718.0  & 23:34:26.9  & $-$54:27:32.5 & 0.53$\pm 0.04$ & 3.41  & $<0.37$ & - & - \\ 
 SCSO~J232215.9-555045.6 & 23:22:17.0 & $-$55:50:07.1 & 0.56$\pm 0.04$  & 2.40  & $0.32_{-0.32}^{+4.13}$  & $29_{-29}^{+2303}$ & $1_{-1}^{+103}$  \\
SCSO~J232247.6-541110.1  & 23:22:53.0  &$-$54:10:54.8 & 0.58$\pm 0.04$  &1.16  & $5.11_{-4.09}^{+5.57}$ & $532_{-495}^{+1286}$ & $138_{-129}^{+333}$ \\
 SCSO~J232211.0-561847.4 & 23:22:13.6  &$-$56:18:35.7 & 0.61$\pm 0.05$  & 5.65  & $0.47_{-0.47}^{+2.61}$ & $48_{-48}^{+1045}$ & $2_{-2}^{+50}$ \\
SCSO~J233731.7-560427.9 &   23:37:30.0 & $-$56:04:01.2 & 0.61$\pm 0.05$ & 3.05  & $2.56_{-2.56}^{+3.87}$ & $276_{-276}^{+1005}$ & $41_{-41}^{+149}$ \\
 SCSO~J234012.6-541907.2  & 23:40:08.8 & $-$54:19:02.9  & 0.62$\pm 0.04$ & 5.23  & $<2.56$  \\
 SCSO~J234004.9-544444.8 & 23:40:02.9 & $-$54:44:21.0  & 0.66$\pm 0.05$ & 4.20  & $6.74_{-3.36}^{+4.45}$   & $843_{-575}^{+1118}$ & $212_{-145}^{+281}$ \\
 SCSO~J232342.3-551915.1 &23:23:45.5 & $-$55:19:08.9 & 0.67$\pm 0.04$ & 2.72  & $<0.51$ &  - & - \\
 SCSO~J232829.7-544255.4 & 23:28:27.5 & $-$54:42:19.3  & 0.68$\pm 0.04$ & 8.34  & $0.59_{-0.59}^{+3.12}$  & $69_{-69}^{+1409}$ & $3_{-3}^{+67}$ \\
 SCSO~J233403.7-555250.7 &   n/a & & 0.71$\pm 0.04$ & 0.88 &  -  &  - & -\\ 
  SCSO~J233951.1-551331.3  & n/a & & 0.73$\pm 0.04$ & 1.3  & -&  - & - \\ 
  SCSO~J233720.2-562115.1  & 23:37:22.4 & $-$56:20:44.8  & 0.75$\pm 0.03$ &0.70  & $5.61_{-4.87}^{+7.24}$& $835_{-806}^{+2488}$ & $151_{-145}^{+450}$  \\ 

\hline
\label{tab2}
\end{tabular}
\end{minipage}
\end{table*}

\section{Conclusions}
There has been longstanding optimism that SZ selection would be among the most favourable ways of detecting clusters for cosmological studies, since simulations show that the SZ detection threshold corresponds very nearly to a threshold in mass.  Despite this, it was not guaranteed that the first SZ experiments could trace mass. In order to demonstrate this we have measured, for the first time, weak lensing masses of clusters detected by their Sunyaev-Zel'dovich signature.  Of the four clusters detected by the SPT and published recently by \citet{SPTSZ}, we have detected three of them, using optical imaging data from the Southern Cosmology Survey. The fourth cluster, at redshift of 0.88, is too distant to be detected with these optical data. By fitting NFW profiles we have established that the published SZ peaks correspond to mass regions, and so we can confirm that the first installment of SZ selected clusters trace the most massive clusters.

We have also presented weak lensing mass estimates for other clusters detected optically in the SCS. As one might expect the published SZ clusters have masses at the upper end of the sample $10^{14}-10^{15}M_\odot$. Furthermore, we have presented analytic predictions for the integrated Compton $Y$ parameter for the all clusters in the sample for future comparison with SZ observations.

\section*{Acknowledgments}

\noindent The authors thank Catherine Heymans for many useful discussions. RNM acknowledges the award of an STFC studentship. FM and JPH acknowledge financial support from the National Science Foundation under the PIRE program (award number OISE-0530095). RJM acknowledges financial support through FP7 grant MIRG-CT-208994 and STFC Advanced Fellowship PP/E006450/1. We acknowledge the support of the European DUEL Research Training Network (MRTN-CT-2006-036133).

\label{lastpage}

\end{document}